\documentstyle[preprint,eqsecnum,aps]{revtex}
\begin{document}
\draft
\tighten
\preprint{UCSBTH-95-5, gr-qc/9504028}
\title{Quantum Probes of Spacetime Singularities}
\author{Gary T. Horowitz\cite{gary} and Donald Marolf\cite{don}}
\address{Physics Department, The University of California,
Santa Barbara, California 93106} \date{April, 1995}
\maketitle

\begin{abstract}

It is shown that there are static spacetimes with timelike curvature
singularities which
appear completely nonsingular when probed with quantum test particles.
Examples include extreme dilatonic black holes and the
fundamental string
solution. In these spacetimes,
the dynamics of
quantum particles is well defined and uniquely determined.

\end{abstract}

\section{Introduction}

In general relativity, a spacetime is considered singular if it is
geodesically incomplete. This is intuitively
reasonable since geodesics
describe the motion of test particles.
Thus, if a spacetime is (timelike)
geodesically
incomplete, then the evolution of some test particle is not defined
after a finite proper time.
The use of geodesic incompleteness
is not only intuitively appealing, it has also been quite
useful in establishing
that large classes of solutions to Einstein's equations are singular.

There has been extensive debate over whether these
singularities in general relativity will be
``smoothed out' in quantum gravity. Various
model systems have been quantized with inconclusive results (see
\cite{L,GD} for some classic treatments).
As a first step toward understanding
the relation between quantum theory
and singularities, we consider the motion
of a quantum test particle in
a classical singular spacetime.
We will see that there are static spacetimes
with timelike singularities in
which a quantum test particle is completely
well behaved for all time.  Even more significantly, these
singularities do not introduce any new ambiguities or
require additional boundary conditions in the
definition of the quantum particle.  The dynamics is uniquely
defined {\it by the spacetime}, just as on a non-singular background.

Thus, even though these spacetimes appear
singular when probed with classical test
particles, they are nonsingular
when the test particles are treated quantum
mechanically.  Roughly speaking,
the reason for the difference is that
these spacetimes produce an effective
repulsive barrier which shields their classical singularity, and
quantum wave packets simply bounce off
this barrier. From this viewpoint,
geodesics correspond to the geometric optics limit
of infinite frequency
waves.  Only in this unphysical limit is the  singularity reached.

Another motivation for studying the motion of quantum test
particles in a classical spacetime comes from string theory.
Classical solutions
to string theory are associated with two dimensional
conformal field theories.
These theories describe the motion of quantum
test strings  in a background
classical geometry. A solution to string theory
is singular if there does not
exist a well defined evolution for these quantum
test strings. Since a string
consists of an infinite number of modes which
represent particles of increasing
mass and spin, studying the behavior of
a single quantum test particle will
give  a preliminary indication of the behavior of a test string.
(Unfortunately, these results will not be conclusive
since even if the quantum particle
is singular, there may
exist an equivalent ``dual" description of the solution
in string theory
which is nonsingular\cite{dual}.)

If one wants to investigate quantum probes of singularities,
one needs a condition in the
quantum theory of the test particle which
determines
whether or not it is singular.
The general definition of a singularity
in a quantum theory is still controversial.
Some people have suggested looking at the expectation value
of certain `physical operators' to see whether they diverge.
The notion of a singularity that we will study is somewhat different.
We will be interested in particular in the analogue of a timelike
singularity.  As such, we will say that a system is {\it non}singular when
the evolution of any state is {\it uniquely} defined for all
time.  If this is not
the case, then there is some loss of predictability and we will
say that the system is singular.

To illustrate this, consider nonrelativistic
quantum mechanics on a bounded interval.  Note that this system
is classically singular as the associated `spacetime' is geodesically
incomplete.
One can initially define the Hamiltonian $H$ to be the Laplacian acting
on wave functions that vanish smoothly at the boundary.
This operator is symmetric, but not yet
self-adjoint. There are in fact many so-called extensions of this
operator (given by defining $H$ to act as the Laplacian on a slightly
larger domain) which are self-adjoint and which correspond to the
different boundary conditions which might be imposed at the edges.
One of these extensions must be chosen
in order to evolve quantum states.
This is directly analogous to a typical
timelike singularity in classical
general relativity, as in this case
one must make a choice of
boundary conditions at the singularity.
In both cases, the evolution is not unique until extra information is
specified. One can imagine another
type of singularity in nonrelativistic
quantum mechanics which is more analogous to a
spacelike singularity than
a timelike one. This occurs if $H$ is time dependent and
is self-adjoint for $t<t_0$ but fails to be self-adjoint
(or even fails to exist) at $t=t_0$.
Here, however, we will concentrate on
the case of timelike singularities.

The hydrogen atom is the prime example of a singular classical
theory which
is nonsingular quantum mechanically.
There is a direct analog of this for singular geometries. Consider
the (non-relativistic) quantum mechanics of a free particle moving
on an $n+1$ dimensional
Reimannian manifold
$M, g$. The Hilbert space consists of square integrable
functions on $M$ with
the measure given by the proper volume element. The Hamiltonian is
proportional to the Laplacian on the manifold. It is known that if
the metric $g$ is geodesically complete, then the Laplacian has a
unique self-adjoint extension  \cite{Lap} (operators for which this is the
case are called
essentially self-adjoint). This means that if the space is
classically nonsingular, then
it is nonsingular quantum mechanically
as well. We are interested in
determining whether the metric can be geodesically incomplete
and still have
a unique self-adjoint Laplacian.

It is easy to see that the answer is yes.
Consider a spherically symmetric metric
\begin{equation}
\label{ssqm}
{ds^2 = dr^2 + R^2(r) d\Omega_n  }
\end{equation}
where $d\Omega_n$ is the standard metric on the $n$-sphere.
We first take the domain of the Laplacian to consist of
smooth functions
with compact support away from the origin.  The key
question is whether the
resulting operator is essentially self-adjoint.
A sufficient condition for this to be the case is to
consider solutions to $D^2 \psi \pm i \psi =0 $
and show that such
solutions are not square integrable \cite{R}.
Using separation of variables,
$\psi = f(r) Y({\rm angles})$, we obtain the radial equation
\begin{equation}\label{radial}
{f'' + {n R' \over R} f' - {c\over R^2} f \pm if =0}
\end{equation}
where a prime denotes the derivative with respect to $r$ and
$c \ge 0$ is an eigenvalue of (minus) the Laplacian on the
$n$-sphere.
Essential self-adjointness is in fact
equivalent \cite{R} to the statement that, for each $c$
and each choice of
$\pm i$, there is one solution to
(\ref{radial}) which fails to be square integrable near the origin.
It suffices to consider the case $c=0$ since increasing
$c$ increases the divergence of one solution at $r=0$.
Near the origin, if $R=r^p$, then the two solutions are
$f=r^\alpha$ where $\alpha=0$ or $\alpha = 1-np$ (note that the
$\pm if$ term is negligible near $r=0$). If $p\ge 3/n$,
the latter solution fails to
be square integrable with respect to the proper volume element
$r^{pn}dr d\mu$,  where $d\mu$ is the volume element on the unit
$n$-sphere. We conclude that any metric of the form
(\ref{ssqm}) which behaves like $R(r) = r^p$
with $p \ge 3/n$ near the origin
is nonsingular in quantum mechanics.
Of course, the metric (\ref{ssqm}) is
geodesically
incomplete unless $p=1$.\footnote{This
argument seems to imply that the Laplacian
in three dimensional Euclidean space is not essentially self-adjoint.
This is indeed the case if one initially defines
the operator only away from
the origin and is consistent with the fact that ${\bf R}^3- \{0\}$ is
geodesically incomplete. However when the space
is regular at the origin
(and only in this
case) one can require $D^2 \psi \pm i\psi =0$ at the origin as
well.  This
removes the ambiguity.} So  there is a large class of
geometries which
are singular classically, but not quantum mechanically. They are
geometric analogues of the hydrogen atom.

\section{Static Spacetimes}

\subsection{General condition for quantum regularity}
\label{gc}

For a static, globally hyperbolic spacetime, there is a well
defined quantum
theory for a single relativistic particle (see, for example,
\cite{part}). We will show that for
certain static spacetimes with timelike
singularities, this is still the case.
We will consider
a relativistic particle with mass $m \ge 0$, which is described
quantum mechanically by a positive frequency
solution to the wave equation of mass $m$.
Some time ago, Wald discussed solutions to
the (massless) wave equation in the presence
of singularities \cite{wald}. Our discussion
will be based on his approach.

Consider a static spacetime with timelike Killing field
$\xi^\mu$. Let
$t$ denote the Killing parameter, and $\Sigma$ denote a static slice.
The wave equation $(\nabla^\mu \nabla_\mu - m^2) \psi =0$ can be
rewritten in the form
\begin{equation}\label{weq}
{{\partial^2 \psi \over \partial t^2} = V D^i(VD_i \psi) } - V^2m^2 \psi
\end{equation}
where $V^2 = -\xi^\mu \xi_\mu$,  and
$D_a$ is the spatial covariant derivative on $\Sigma$.
Let $A$ denote (minus) the
operator on the right hand side
\begin{equation}\label{spaop}
{ A \equiv -VD^i(VD_i) + V^2 m^2. }
\end{equation}
Consider the Hilbert space ${\cal H}$ of square integrable
functions on $\Sigma$
with the inner product $V^{-1}$ times the proper volume
element. If we
initially define the domain of $A$ to be smooth functions of
compact support
on $\Sigma$, then since $V^2m^2 \ge 0$,
 $A$ is a positive symmetric operator.  (Recall that in
general relativity, the `singular points' are not included
as part of $\Sigma$.)  We note that $A$ of \ref{spaop} is also
a {\it real} differential operator so that its deficiency
indices\cite{R} are always equal and self-adjoint extensions always
exist.
The key question is whether such an extension is unique.
If it is, then this extension
$A_E$ is
always positive definite and we may define
its positive self-adjoint square root.
Then the wave function for a free relativistic
particle satisfies
\begin{equation}\label{frpeq}{i {d \psi \over dt} =
(A_E)^{1/2}\psi }
\end{equation}
with solution
\begin{equation}\label{soln}{ \psi(t) = \exp{\
[-it(A_E)^{1/2}\ ]} \psi(0). }
\end{equation}
The right hand side is well defined using standard properties of
self-adjoint operators.  If there is more than one
self-adjoint extension
of $A$, then (\ref{frpeq}) and (\ref{soln}) are ambiguous.
This is our criterion for
calling the quantum theory  singular.

Wald \cite{wald}
considered the second order wave equation. He did not require the
operator $A$ to be essentially self-adjoint, but instead picked an
arbitrary positive definite
self-adjoint extension and studied the resulting solution. He showed
that it agreed with the usual Cauchy evolution
inside the domain of dependence of the initial surface.

\subsection{Examples}

In this section we will consider some examples of static
solutions that have
recently been discussed
in the literature. All of these solutions are geodesically
incomplete, and we wish to determine whether they are singular
when probed by
quantum test particles. We first consider a general static,
spherically symmetric
metric in $n+2$ dimensions
\begin{equation}\label{ssfd}
{ ds^2 = -V^2 dt^2 + V^{-2} dr^2 + R^2 d\Omega_n  }
\end{equation}
where $V$ and $R$ are functions of $r$ only.
As discussed above, the crucial question is whether
the spatial operator $A$ (\ref{spaop}) is
essentially self-adjoint. Consider the
equation $A \psi \pm i \psi = 0$.
Separating variables $\psi = f(r) Y({\rm angles})$
leads to the following radial equation for $f$
\begin{equation}
\label{radeq}
{ f'' + {(V^{2} R^n)' \over V^{2} R^n } f' -{c \over V^{2} R^2}
f - {{m^2} \over {V^2}}f  \pm i {f \over V^4} =0}.
\end{equation}
The operator
$A$ will be essentially self-adjoint if one of the two
solutions to this
equation (for each $c$ and each sign of the imaginary term)
fails to be square integrable with respect to the measure
$R^n V^{-2}$ near $r=0$.

Suppose that, for $m=0$, one solution of (\ref{radeq}) fails to be
square integrable near the origin.  Then, because $m^2 \ge 0$ and
$V^2 \ge 0$, the addition of the term $- {{m^2 f} \over {V^2}}$ acts
like a repulsive potential in quantum mechanics.  That is, it will
increase the rate at which the larger solution diverges at the origin
while driving the other more quickly to zero.  It
follows that if $A$ is
essentially self-adjoint for $m=0$, it is essentially self-adjoint
for all $m \ge 0$ as well.  Thus, we need only consider the massless
case below.

The metric (\ref{ssfd}) can have a null singularity instead
of a timelike one.
The difference is seen as follows. Define a new radial coordinate
$dr_* = dr/V^2$, so that
radial null geodesics follow curves of constant $t \pm r_*$.
If the singularity is at a finite value of $r_*$, then it is timelike.
But if it is at $r_* = -\infty$, then it is null.  (A Penrose
diagram of the resulting spacetime would resemble the region $r>2M$
of the Schwarzschild solution, with
a singularity along the horizon $r=2M$.) If the
singularity is null, the spacetime is globally hyperbolic. It follows
immediately that the operator $A$ must be essentially
self-adjoint. This
is because, if there were more than one self-adjoint
extension, there would
be two distinct evolutions of initial data to the wave equation,
of the form described by Wald \cite{wald}.
But these solutions must agree with ordinary Cauchy evolution
(which is unique),
so all self-adjoint extensions of $A$ must agree. We will therefore
consider only spacetimes with timelike singularities.

Consider first the (four dimensional) negative mass
Schwarzschild solution. It is easy to verify that both solutions to
(\ref{radeq})
are locally normalizable near $r=0$. Thus,
it remains singular even
when probed with quantum test particles. This is fortunate,
since if the
negative mass Schwarzschild solution was nonsingular in some theory,
then that theory would probably not have a stable ground
state \cite{homy}.
One can also verify that the Reissner-Nordstr\"om
solution remains singular for all values of the charge to
mass ratio $Q/M$.
It is interesting to note that the $M < 0$
Schwarzschild solution is timelike
geodesically complete.  As a result, a massive relativistic classical
particle in this
spacetime is nonsingular while the corresponding quantum
theory is singular.  We thus have a counterexample to Wheeler's
`rule of unanimity' \cite{Un}\footnote{A similar, but simpler,
counterexample is given by the Hamiltonian $p^2 + 1/x$ for a
nonrelativistic particle on the half line $x>0$.}.

We now consider four dimensional,
charged dilatonic black holes. They are extrema
of the following action
\begin{equation}\label{action}{
S=\int d^4x
{\sqrt {-g}}\left[R-2(\nabla\phi)^2-e^{-2a\phi}F^2\right]}
\end{equation}
where $\phi$ is the dilaton, $F$ is the Maxwell field, and $a$ is
a constant
which governs the strength of the dilaton coupling.
For $a = \sqrt 3$,
this action is equivalent to Kaluza-Klein theory.
In other words, given
an extremum of (\ref{action}) with this value of $a$,
one can reconstruct a
solution of the five dimensional vacuum Einstein
equation. The charged
black hole solution to this theory (for general $a$) is given by
a metric of the form (\ref{ssfd}) with \cite{gima}
\begin{equation}\label{dcbh}{ V^2=\left(1-{r_+\over r}\right)
 \left(1-{r_-\over r}\right)^{{(1-a^2)\over (1+a^2)}},\qquad
  R^2=r^2\left(1-{r_-\over r}\right)^{2a^2\over (1+a^2)} }
\end{equation}
Notice the product of these two quantities is independent of $a$ and
is simply
\begin{equation}\label{prod}{ V^2 R^2 = (r-r_+)(r-r_-) }.
\end{equation}
For $r_+ > r_-$ and $a \ne 0$,
this metric describes a black hole with an event horizon
at $r=r_+$ and a singularity at $r=r_-$. (For the special case $a=0$,
$r=r_-$ denotes the inner Cauchy horizon of
the Reissner-Nordstr\"om solution
which is nonsingular.)
The extremal limit $r_+ = r_-$ describes a globally static spacetime
with a curvature singularity at $r=r_+$.
This singularity is null for $a\le 1$,
but is timelike for $a >1$.

The operator $A$ (\ref{spaop})
must be essentially self-adjoint for  $r_+ = r_-$ and
$a\le 1$ since the
singularity is null. We wish to investigate whether
this continues to be the
case for $a >1$.  To begin,  let $\rho = r-r_+$, so $V^2 R^2 = \rho^2$.
Since $a >1$,  $V^2 > \rho$ so that the imaginary
term in (\ref{radeq}) may be ignored.
Then one solution to (\ref{radeq}) behaves like $f=\rho^\alpha$
with $\alpha \le -1$ near the singularity $\rho=0$.
The least divergent solution, $\alpha = -1$, corresponds to
the $S$-wave, $c=0$. From (\ref{dcbh})
we see that this solution has norm
\begin{equation}\label{norm}{ <f|f> = \int \rho^{-4/(1+a^2)} d\rho}
\end{equation}
which diverges near $\rho = 0$ for $a^2 \le 3$.
Thus, extremal dilaton black holes with $1<a^2 \le 3$
are examples of static spacetimes with timelike
singularities for which
quantum test particles are well behaved.
The fact that the solutions (\ref{dcbh})
have infinite repulsive barriers when $a^2 >1$ was noticed earlier
by Holzhey and
Wilczek \cite{HW}.
However, their analysis did not distinguish between $a^2$ greater
than three and less than three.
We now see that for $a^2 > 3$ quantum mechanics does not
exclude the solution that grows near $\rho = 0$.

Notice that extreme
Kaluza-Klein black holes are included in the class
of solutions which are
quantum mechanically nonsingular.
One might wonder if this is related to the fact that
the {\it five} dimensional
metric for an extreme magnetically charged black hole does not have a
curvature singularity. The answer is clearly no. First,
our analysis applies to
both electric and magnetically charged solutions (since the metric
(\ref{dcbh}) is the same) and the electrically
charged solution remains
singular in five dimensions. Second,
if one dimensionally reduces the $4+m$ dimensional
Einstein action to four dimensions one can obtain
the action (\ref{action})
with $a=\sqrt{(m+2)/m}$ \cite{gima}. So all of these
extremal Kaluza-Klein
black holes are nonsingular quantum mechanically, even though  most
have curvature singularities
in $4+m$ (as well as four) dimensions.

As another example, we consider the fundamental string
solution discovered
by Dabholkar et al. \cite{FS}.
This was originally found as a solution to the low
energy string action
\begin{equation}\label{stact}
{ S = \int d^D x \sqrt{-g} e^{-2\phi} [ R + 4(\nabla \phi)^2 -
{1\over 12} H^2]}
\end{equation}
(where $D$ is the spacetime dimension and $H$ is the three form)
but was later shown to be an exact solution
to string theory \cite{exact}.
The metric is given by
\begin{equation}\label{fssol}{ ds^2 = V^2
(-dt^2 + dz^2) + dx_i dx^i}
\end{equation}
\begin{equation}\label{fss}{ V^{-2}
={ 1 + {M \over r^{D-4}} }   \ , }
\end{equation}
where $r^2 = x_i x^i$.
This solution
describes the field outside of a straight
fundamental string located at
$r=0,$ which is a curvature singularity.
This singularity is null for $D\ge 6$
but is timelike for $D=5$.  Thus, $A$ must be
essentially self-adjoint
for $D \ge 6$. By performing an analysis similar
to that above, one can
show that $A$ remains essentially self-adjoint when $D=5$. So this
provides another example of a classically singular spacetime which is
nonsingular quantum mechanically.

However, this result is not directly applicable to singularities in
string theory since we have
not included the  effect of the dilaton on the test particle.
Recall that the
lowest mode of a (bosonic) string is the tachyon which is
coupled to the dilaton via
\begin{equation}\label{dilan}
{  S= \int d^D x \sqrt{-g} e^{-2\phi}[(\nabla \psi)^2 + m^2 \psi^2]}.
\end{equation}
For a static spacetime, the wavefunctions of the tachyon modes
satisfy the equation of motion
\begin{equation}\label{diltac}
{ {\partial^2 \psi \over \partial t^2} = - \tilde A \psi}
\end{equation}
where
\begin{equation}\label{defa}
{ \tilde A =  - V e^{2\phi} D_i[Ve^{-2\phi} D^i \psi]  + m^2 V^2}
\end{equation}
and the notation is the same as in (\ref{spaop}).  Since $m^2 < 0$
for the tachyon, we must keep the mass term for now.
This operator is symmetric
with respect to an $L^2$ inner product with measure equal to the
proper volume element divided by $V e^{2\phi}$.

We now show that
$\tilde A$ is essentially self-adjoint for the $D=5$
fundamental string (\ref{fssol}).
The dilaton for this solution is given by $e^{\phi} = V$. After
separating variables
$\psi = f(r) e^{ikz} Y({\rm angles})$, the equation
$\tilde A \psi \pm i\psi = 0$
yields the following radial equation for $f$:
\begin{equation}\label{fsreq}
{f'' + {2 \over r} f' - \left ( {{k^2} \over V^2}
+ {c \over r^2 } \right ) f - m^2 f  \pm i{f \over V^2} =0   }
\end{equation}
where $c\ge 0$ is again an eigenvalue of (minus) the
Laplacian on the sphere.
Since $V^2 = r/M$ near $r=0$, we see that the $k^2$ term,
the mass term,
and the imaginary term
are all negligible near the origin.
Thus, one solution in this region
is $f = r^\alpha$ where $\alpha \le -1$. This solution
always has infinite norm
near $r=0$ since the appropriate inner
product is
\begin{equation}\label{fsipd}{ <f|f> = \int {|f|^2
V r^2 dr\over V e^{2\phi}}
	 = \int |f|^2 M r dr }.
\end{equation}
Therefore, even when the coupling
to the dilaton is included, the singularity in the
fundamental string
does not prevent unique evolution of the tachyon. This suggests that
other modes of the string will similarly have unique evolution, but
the effect of spin needs to be investigated.

Another exact solution to string theory is an orbifold,
which is constructed
by starting with flat Euclidean space and identifying
points under the
action of a discrete group. If the group has
fixed points, then the quotient
is geodesically incomplete. Nevertheless, it is believed
that string theory is well
behaved on these backgrounds \cite{orb}.
 From our discussion in the introduction, it
is clear that for a two dimensional orbifold
(which is a cone), the operator
governing evolution of a scalar test particle (or the tachyon)
is {\it not} essentially
self-adjoint. This suggests that the
propagation of test strings is  also not well defined
without further specification of boundary
conditions at the singularity.
This is not a problem in dimensions greater than three, so the most
commonly discussed case of a six
dimensional orbifold is nonsingular for quantum test particles.

\subsection{Scattering}

The evolution defined by $A_E^{1/2}$ has all of the nice
properties of familiar quantum mechanical systems.  By construction,
the evolution is unitary and the energy $A_E^{1/2}$ is conserved.
However, we have not yet ruled out the possibility that an incoming
wave packet might remain localized near the singularity, resulting
in a nonunitary S-matrix. Indeed, we expect this will happen whenever
the singularity is null, since the wave then takes an infinite
(coordinate) time to reach
the singularity.
However,
we now show that, at least for highly symmetric cases,
this cannot occur for timelike singularities.
For such cases, the S-matrix
is unitary.

Consider a spherically symmetric metric
of the form (\ref{ssfd}) with a timelike singularity at the origin.
As usual, spherical
symmetry and time independence imply that energy
and angular momentum are
conserved in the scattering so that  we can confine our
attention to the radial eigenfunction equation.
Since any eigenstate of
$A_E^{1/2}$ is also an eigenstate of $A_E$,  it is in fact
sufficient to study wavefunctions $f$ that solve

\begin{equation}\label{Eeq}{f'' + {{(V^2R^n)'}  \over {V^2R^n}} f' -
{c \over V^2 {R^2}}f  -{m^2 \over V^2}f + {{Ef} \over V^4} = 0.}
\end{equation}

Let $R=r^p$ and $V=r^q$ near the origin, and
consider first the case $c=m=0$.
Since the singularity is timelike, $q < 1/2$.
Thus, the term $Ef/V^4$
is negligible near $r=0$, and the two solutions to (\ref{Eeq})
take the form  $f =
r^{\alpha}$  with $\alpha = 0, 1-2q-np$. By our previous discussion, the
condition that the classical singularity not affect
quantum test particles
is that the solution $r^{1-2q-np}$ must not be square integrable
near $r=0$ with respect to the measure $R^n V^{-2} dr$.
Since this measure is $r^{np-2q}$ near the origin,
the condition that the singularity be
timelike ($q<1/2$) guarantees that the other solution $r^0$
is always square integrable.
If $c$ and $m^2$ are nonzero, the above equation is modified by the
addition of a repulsive potential (assuming nontachyonic particles)
which increases the
divergence of the more singular solution and
forces the less singular solution
to vanish more quickly.
Thus, for any $c \ge 0$ and $m^2 \ge 0$,
there is exactly one allowable solution of (\ref{Eeq}).
It is real, with equal incoming and outgoing
flux.  Thus, the $S$-matrix is unitary.
A similar argument establishes unitarity for
nontachyonic particles in
any cylindrically symmetric spacetime. For the special case
of the $D=5$ fundamental string
solution (\ref{fssol}), one can verify that tachyon scattering is
also unitary.

\section{Extensions}

In the previous section, we considered
only the propagation of quantum test particles
on a static (time-independent) background.  Any extension to more
general
cases will clearly require a change of outlook, if not of techniques.
Indeed, for a general time-dependent background there
is no consistent
quantum theory of a single free particle in the usual sense and
the only appropriate
description is in terms of quantum field theory.
Since linear quantum
field theory is defined by the solutions of classical
field theory,  the essential step is to study
the evolution of classical test fields on a singular background.

This may not be as difficult as it sounds.  As
described in \cite{wald}, techniques similar to those applied
here can be used to define classical field evolution in static
singular spacetimes.
Given a scalar field $\phi$ satisfying a wave equation of the form
\begin{equation}
\label{swave}
{{{\partial^2} \over {\partial t^2}} \phi = -A \phi}
\end{equation}
with $A$ a symmetric operator on the Hilbert space ${\cal H}$
of section II A and any self-adjoint extension
$A_E$ of $A$, the field
\begin{equation}\label{was}
{\phi(t) = \cos[A_E^{1/2}t] \phi(0) + A_E^{-1/2} \sin[A_E^{1/2}t]
{\dot{\phi}}(0)}
\end{equation}
is the  unique solution of ${{\partial ^2} \over {\partial t^2}}
\phi = -A_E \phi$ (which takes the value $\phi(0)$ at $t=0$
and has time derivative $\dot{\phi}(0)$ at $t=0$)
and also satisfies (\ref{swave}) in any
hyperbolic domain.  Thus, when $A$ is essentially
self-adjoint, there is
a unique solution of this form and no boundary conditions need be
imposed.

What  about the general  nonstatic case? It is not difficult to
make the first steps.  By reformulating the general
wave equation in the first
order form
\begin{equation}\label{fof}
{ {\partial \over {\partial t}} \Biggl[ {\phi(t) \atop
\dot{\phi}(t)} \Biggr] = \Biggl[ {0 \atop -A(t)}
{1 \atop iB(t)} \Biggr] \Biggl[
{\phi(t) \atop \dot{\phi}(t)} \Biggr]},
\end{equation}
it is clear that our task is to define the path ordered exponential
\begin{equation}\label{fsol}
{ \Biggl[ {{\phi(t)}
\atop  {\dot{\phi}(t)}}
\Biggr] = {\cal P}\exp\Biggl(\int_0^t \Biggl[ {0 \atop -A}
{1 \atop {iB}} \Biggr]\Biggr)
\Biggl[ {{\phi(0)} \atop {\dot{\phi}(0)}}\Biggr].}
\end{equation}
As before, we will need to work in certain Hilbert spaces, and the
choices
\begin{equation}\label{htdef}
{\Biggl( \Biggl[ {{\phi_1} \atop {\dot{\phi}_1}} \Biggr],
\Biggl[ {{\phi_2} \atop {\dot{\phi}_2}} \Biggr] \Biggr)_{{\cal H}_t}
 = \int_{\Sigma_t} (\phi_1^*\phi_2 + \dot{\phi}_1^*\dot{\phi}_2)
\sqrt{-g} g^{tt} d^{n-1}\Sigma_t}
\end{equation}
are natural.
Note that in the static case $\sqrt{-g} g^{tt}  d^{n-1}\Sigma_t$
is $V^{-1}$ times
the proper volume element on $\Sigma_t$, so that this is a
straightforward generalization of ${\cal H}$ from
section \ref{gc}.
In the special case of a static spacetime, the path ordered
exponential is well-defined and gives the solution (\ref{was}).
The case in which the spacetime is stationary (so that ${\cal H}$,
$A$, and $B$ are time independent) and $[A,B] = 0$ is also
straightforward to exponentiate and yields
\begin{equation}\label{sysol}
{\phi(t) = e^{itB/2} \cos\left ({t \over 2} \sqrt{B^2 + 4A}\right ) \phi(0)
+ e^{itB/2} {2 \over {\sqrt{B^2 + 4A}}}
\sin\left ({t\over 2} \sqrt{B^2 +4A}\right )
[\dot{\phi}(0) - i {B \over 2} \phi(0)]}.
\end{equation}

While the general time dependent case remains to be investigated,
we mention that the following two results can be derived by
elementary methods.  First, by using an `interaction picture,'
it is readily shown that if $A(t)$ and $B(t)$
differ from the operators
associated with either of the solutions (\ref{was}) or (\ref{sysol})
by an appropriately bounded perturbation\footnote{Unfortunately,
since perturbations of wave operators are polynomial
differential operators, they are not bounded, so that this
case is not of direct physical interest.}, then there is a unique
(and well-defined) solution of the form (\ref{fsol}).
Also, by assuming that
a solution of the form (\ref{fsol}) is well-defined, it is
readily shown (using much the same method as \cite{wald})
to agree with the solution of the
usual wave equation in any hyperbolic
domain.  Such a solution also
conserves the Klein-Gordon inner product over the
entire spacetime.

Whether such ideas can be developed further is an interesting
question for future research.  Also of interest
would be a search for
corresponding results for higher spin fields. This
would be an important
step toward extending these results from test
particles to test strings.
Since the (four dimensional)
Maxwell equations are conformally invariant,
one can construct examples of singular spacetimes in which
Maxwell fields are well behaved but scalar fields are not.
If it is found that a large class of fields have  nonsingular
evolution on some singular background, then
such a spacetime need not be seen as a threat to cosmic censorship.
Instead of being shielded by a horizon, the timelike singularity
would be shielded by the effective repulsive
barrier that it presents to
wave propagation.

\acknowledgments
The authors would like to thank John Baez, Patrick Brady, and Alan
Rendall for a number of useful discussions.  This work was supported
by NSF grant PHY90-08502.

\end{document}